\documentclass[fleqn,usenatbib,useAMS]{mnras}

\usepackage{graphicx}	% Including figure files
\usepackage{amsmath}	% Advanced maths commands
\usepackage{multicol}        % Multi-column entries in tables
\usepackage{bm}		% Bold maths symbols, including upright Greek
\usepackage{pdflscape}	% Landscape pages

\newcommand{\cwr}[1]{{\color{black}{#1}}}
\usepackage[modulo]{lineno}
\usepackage{siunitx}
\usepackage[bottom]{footmisc}
\usepackage[T1]{fontenc}
\usepackage{ae,aecompl}

\usepackage{newtxtext,newtxmath}
\title[Rubin DP1 Photo-$z$]{Photometric Redshift Estimation for  Rubin Observatory Data Preview 1 with Redshift Assessment Infrastructure Layers (RAIL)}

\author[]{T. Zhang$^{1,\dag}$,
E. Charles$^{2,3}$,
J.F. Crenshaw$^{4,5}$,
S.J. Schmidt$^{6}$,
P. Adari$^{2,7}$,
J. Gschwend$^{8}$,
S. Mau$^{2,3}$,
\newauthor B. Andrews$^{1}$,
E. Aubourg$^{9}$,
Y. Bains$^{10}$,
K. Bechtol$^{11}$,
A. Boucaud$^{9}$,
D. Boutigny$^{12}$,
P. Burchat$^{2,3}$,
\newauthor J. Chevalier$^{13}$,
J. Chiang$^{2}$,
H.-F. Chiang$^{2}$,
D. Clowe$^{14}$,
J. Cohen-Tanugi$^{15}$,
C. Combet$^{16}$,
A. Connolly$^{4}$,
\newauthor S. Dagoret-Campagne$^{13}$,
P.N. Daly$^{17}$,
F. Daruich$^{18}$,
G. Daubard$^{19}$,
J. De Vicente$^{20}$,
H. Drass$^{18}$,
\newauthor K. Fanning$^{2}$,
E. Gawiser$^{21}$,
M. Graham$^{4,5}$,
L.P. Guy$^{22}$,
Q. Hang$^{23}$,
P. Ingraham$^{18}$,
O. Ilbert$^{24}$,
\newauthor M. Jarvis$^{25}$,
M.J. Jee$^{26,6}$,
T. Jenness$^{18}$,
A. Johnson$^{2}$,
C. Juramy-Gilles$^{19}$,
S.M. Kahn$^{27}$,
\newauthor J.B. Kalmbach$^{2,3,5}$,
Y. Kang$^{18,2}$,
A. Kannawadi$^{28}$,
L.S. Kelvin$^{29}$,
S. Liang$^{2}$,
O. Lynn$^{30}$,
N.B. Lust$^{29}$,
\newauthor M. Lutfi$^{18}$,
A. Malz$^{30}$,
R. Mandelbaum$^{30}$,
S. Marshall$^{2}$,
J. Meyers$^{2}$,
M. Migliore$^{9}$,
M. Moniez$^{13}$,
\newauthor J. Neveu$^{19}$,
J.A. Newman$^{1}$,
E. Nourbakhsh$^{29}$,
D. Oldag$^{4,5}$,
H. Park$^{28}$,
S. Pelesky$^{30}$,
\newauthor A.A. Plazas Malagón$^{2,3}$,
B. Quint$^{18}$,
M. Rahman$^{31}$,
A. Rasmussen$^{2}$,
K. Reil$^{2}$,
W. Roby$^{32}$,
A. Roodman$^{2}$,
\newauthor C. Roucelle$^{9}$,
M. Salvato$^{33}$,
B. Sánchez$^{34}$,
D. Sanmartim$^{18}$,
R.H. Schindler$^{3,2}$,
J. Scora$^{31}$,
J. Sebag$^{18}$,
\newauthor N. Sedaghat$^{4}$,
I. Sevilla-Noarbe$^{20}$,
R. Shirley$^{33}$,
A. Shugart$^{18}$,
R. Solomon$^{12}$,
D. Taranu$^{29}$,
G. Thayer$^{2}$,
\newauthor L. Toribio San Cipriano$^{20}$,
E. Urbach$^{35}$,
Y. Utsumi$^{2}$,
W. van Reeven$^{18}$,
A. von der Linden$^{7}$,
C.W. Walter$^{28}$,
\newauthor W.M. Wood-Vasey$^{1}$,
J. Zuntz$^{36}$,
LSST Dark Energy Science Collaboration\\
(Affiliations can be found after the references)}

\date{Accepted XXX. Received YYY; in original form ZZZ}

\pubyear{\the\year{}}

\begin{document}
\label{firstpage}
\pagerange{\pageref{firstpage}--\pageref{lastpage}}

\maketitle

% \begin{center}
% $^{\ast}$\,\texttt{tq.zhang@pitt.edu}
% \end{center}

% 
% Abstract of the paper
\begin{abstract}
We present the first systematic analysis of photometric redshifts (photo-$z$) estimated from the Rubin Observatory Data Preview 1 (DP1) data taken with the Legacy Survey of Space and Time (LSST) Commissioning Camera. Employing the Redshift Assessment Infrastructure Layers (RAIL) framework, we apply eight photo-$z$ algorithms to the DP1 photometry, using deep $ugrizy$ coverage in the Extended Chandra Deep Field South (\texttt{ECDFS}) field and $griz$ data in the \texttt{Rubin\_SV\_38\_7} field. 
In the \texttt{ECDFS} field, we construct a reference catalog from spectroscopic redshift (spec-$z$), grism redshift (grism-$z$), and multiband photo-$z$ for training and validating photo-$z$. Performance metrics of the photo-$z$ are evaluated using spec-$z$s from \texttt{ECDFS} and Dark Energy Spectroscopic Instrument Data Release 1 samples.  { Across the algorithms, we achieve per-galaxy photo-$z$ scatter of {$\sigma_{\rm NMAD} \sim 0.03$} and outlier fractions around 10\% in the 6-band data, with performance degrading at faint magnitudes and $z>1.2$. The overall bias and scatter of our machine-learning based photo-$z$s satisfy the LSST Y1 requirement. } We also use our photo-$z$ to infer the ensemble redshift distribution $n(z)$. 
 {We study the photo-$z$ improvement by including near-infrared photometry from the Euclid mission, and find that Euclid photometry {improves} photo-$z$ at $z>1.2$.} Our results validate the RAIL pipeline for Rubin photo-$z$ production and demonstrate promising initial performance. 
\end{abstract}

% Select between one and six entries from the list of approved keywords.
% Don't make up new ones.

% \footnotetext[1]{\dag～\href{mailto:tq.zhang@pitt.edu}{\rm tq.zhang@pitt.edu}}

\renewcommand\thefootnote{}          % temporarily disable footnote symbols
\footnotetext{\dag~\href{mailto:tq.zhang@pitt.edu}{tq.zhang@pitt.edu}}  % your footnote text
\addtocounter{footnote}{-1}          % keep the counter consistent
\renewcommand\thefootnote{\arabic{footnote}}
%%%%%%%%%%%%%%%%%%%%%%%%%%%%%%%%%%%%%%%%%%%%%%%%%%

%%%%%%%%%%%%%%%%% BODY OF PAPER %%%%%%%%%%%%%%%%%%

% \tianqing{TQ - Change all $z_{spec}$, $z_{true}$ to $z_{\rm ref}$}

\section{Introduction}
\label{sec:intro:0}

% \thanks{$\dag$~\href{mailto:tq.zhang@pitt.edu}{\rm tq.zhang@pitt.edu}.}

The {NSF}–{DOE} Vera C. Rubin Observatory Legacy Survey of Space and Time \citep[LSST;][]{2019ApJ...873..111I} is a 10-year survey that will repeatedly image the southern sky in six optical bands ($ugrizy$).  During its operation LSST will produce an unprecedentedly large dataset of billions of galaxies, providing a statistical foundation to probe the nature of dark energy through multiple cosmological probes, including weak gravitational lensing \citep{Kilbinger:2014cea}, large-scale structure \citep{VanWaerbeke2000}, and supernovae \citep{Scolnic2018,DES_Supernovae2019}.

Rubin Observatory's Data Preview 1 \citep[DP1;][]{RTN-095} provides real observations from the LSST Commissioning Camera \citep[LSSTComCam;][]{10.71929/rubin/2561361} mounted on the 8.4-meter Simonyi Survey Telescope, with multiband imaging across selected fields. These datasets \citep{10.71929/rubin/2570308} are the first real data released by the Rubin Observatory and are a testbed to validate and understand the data processing pipelines.

Photometric redshifts (photo-$z$) with well-controlled systematic biases are essential to measuring and modeling the aforementioned probes, and therefore are critical to achieving the science goals of the LSST Dark Energy Science Collaboration \citep[DESC;][]{DMTN-049, newman_photometric_2022}. 
Precise and accurate photo-$z$ estimates for the galaxies are required to select the galaxies for tracing the large-scale structure, { dividing lens and source galaxies into tomographic bins.} Photo-$z$ is also used to infer the redshift distribution of galaxy ensembles and calibrate standard candles in the absence of spectroscopic coverage.

In this paper, we present the photo-$z$ of galaxies in Rubin Observatory's DP1 dataset, produced using the Redshift Assessment Infrastructure Layers \citep[RAIL;][]{RAIL}. We use a compilation of spectroscopic, grism and deep multiband photo-$z$ galaxies to construct the reference sample; we use the reference sample to train and test machine-learning photo-$z$ algorithms, and calibrate the template-fitting algorithms in RAIL; we evaluate the the photo-$z$ algorithms performance on the test set and validation set, made by cross-matching DP1 galaxies to the  {Dark Energy Spectroscopic Instrument (DESI)} Data Release 1 DR1 galaxies \citep{desi-dr1}; { we study the improvement by including near-infrared (NIR) photometry from the Euclid mission \citep{Euclid2025}}; we produce per-galaxy photo-$z$ for DP1 galaxies that pass certain quality flags. This work is an extension of the initial study of DP1 photo-$z$ by \citet{SITCOMTN-154}.  {The photo-$z$ in this work are essential to other DESC early science projects, e.g., a cluster lensing analysis of the Abell 360 cluster (The LSST Dark Energy Science Collaboration et al. {\it in prep.} ).}

The paper is organized as follows: Section~\ref{sec:data:0} describes the DP1 datasets and reference redshift samples; Section~\ref{sec:method:0} describes the photometric redshift estimation methods and our bookkeeping software; Section~\ref{sec:res:0} presents the photo-$z$ performance results; and Section~\ref{sec:conclu:0} summarizes our findings and outlines future studies.

\begin{figure*}
    \centering
    \includegraphics[width=1.0\linewidth]{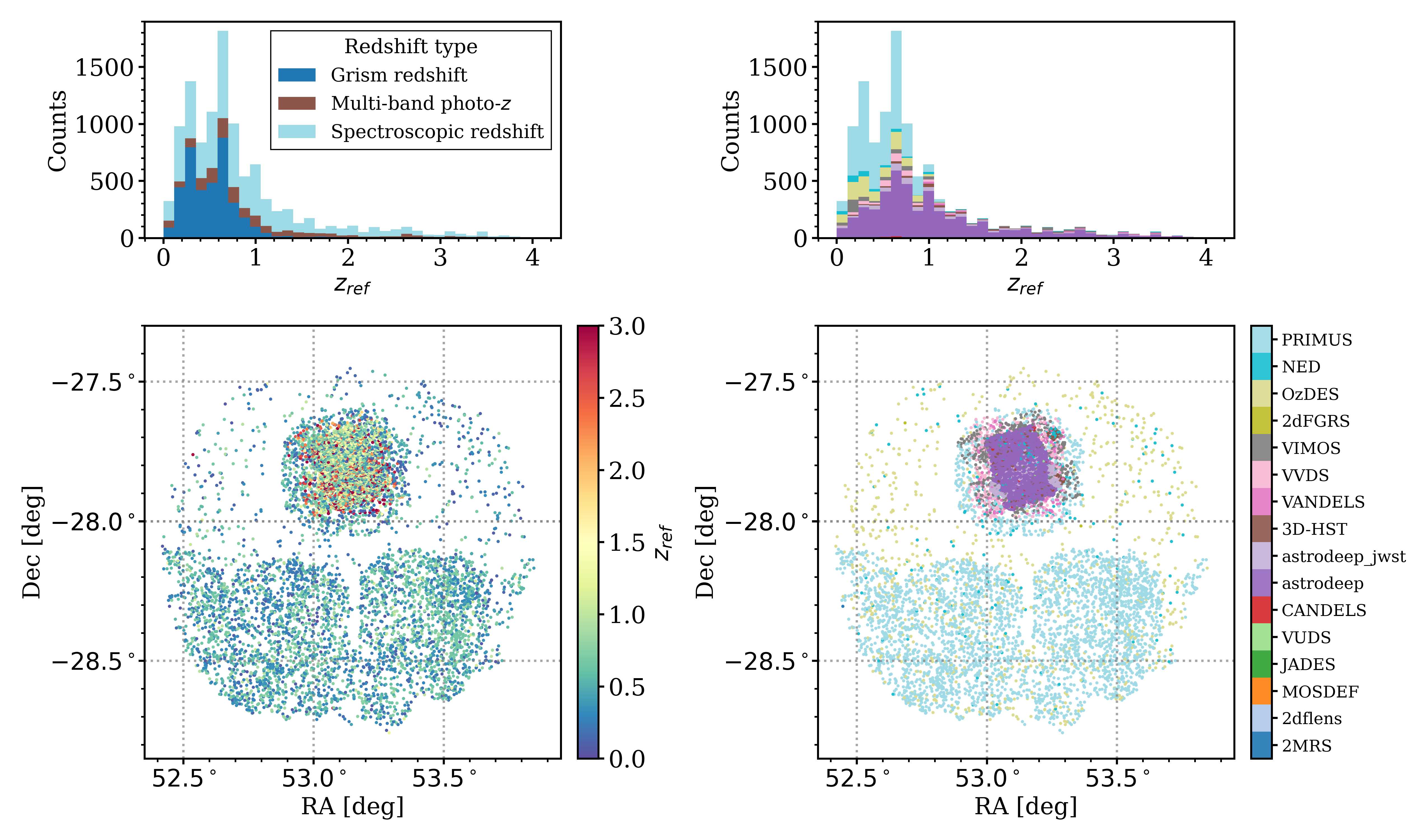}
    \caption{Training set in the \texttt{ECDFS} field. Top left: redshift distribution by methods to obtain redshift. Top right: redshift distribution by surveys. Bottom left: Scatter plot of the \texttt{ECDFS} reference catalog color-coded by redshift. Bottom right: scatter plot of the \texttt{ECDFS} reference catalog color-coded by the survey name.    { The 1-D distributions are normalized. Our reference galaxy catalog is constructed by a wide range of redshifts, and spans a wide range of redshifts.  }}
    \label{fig:train_info}
\end{figure*}

\section{Data}
\label{sec:data:0}

\subsection{Rubin DP1}
\label{sec:data:dp1}

The Rubin Data Preview 1 (DP1) includes $\sim 15 {\rm deg}^2$ of multiband optical imaging $ugrizy$ across six selected fields \citep{RTN-095}, including the Extended Chandra Deep Field South (\texttt{ECDFS}), Euclid Deep Field South (\texttt{EDFS}), a low galactic latitude field \texttt{Rubin\_SV\_95\_-25} (\texttt{SV\_95\_-25}) fields, and a low ecliptic latitude field \texttt{Rubin\_SV\_38\_7} (\texttt{SV\_38\_7}) field with $griz$ photometry. The photometric catalogs consist magnitudes using a variety of flux measurement methods, including the 1\arcsec\ Gaussian Aperture (Gaap1p0) fluxes, the 3\arcsec\ Gaussian Aperture (Gaap3p0) fluxes, the CModel fluxes,  {the Kron fluxes \citep{kron1980}}, the PSF aperture fluxes, and the S\'ersic aperture fluxes from the DP1 coadd object catalog \citep{10.71929/rubin/2570313}.   { We estimate photo-$z$ in these four fields with the Gaap1p0 photometry. } \cwr{The stacked PDF of our photo-$z$ is shown in Appendix~\ref{ap:stacked_pz}. }

To ensure the quality photometry necessary for photometric redshift estimation, we include only galaxies with $i$-band PSF flux signal-to-noise (SNR) ratio exceeding 5. Furthermore, we selected extended sources by requiring $\mathtt{g\_extendedness} > 0.5$ and $\mathtt{r\_extendedness} > 0.5$, which selects against point-source objects. 
 {The $5\sigma$ limiting magnitudes of Gaap1p0 of the \texttt{ECDFS} field are $[26.4, 27.8, 27.1, 26.7, 25.8, 24.6]$.} 
% \tianqing{Describe the other PSF flag too.} 

In the ECDFS, EDFS, and \texttt{SV\_95\_-25} fields, this selection yielded galaxies with six-band photometry, resulting in a sample of approximately 375{,}000 galaxies. In the \texttt{SV\_38\_7} field, where observations are limited to four bands ($griz$), the same selection produced a sample of about 169{,}000 galaxies. For reference, the full DP1 object catalog contains roughly 2 million objects before applying these selection criteria. 

Milky Way dust extinction reduces the flux disproportionately more at bluer wavelengths, and needs to be corrected (dereddened).  We use the extinction maps of \citet{SFD} to get $ E(B-V)(\alpha, \delta)$ values, and use attenuation coefficient $a_{b, V} = [4.81, 3.64, 2.70, 2.06, 1.58, 1.31]$ in the $u, g, r, i, z, y$ bands. The dereddened magnitudes are 
\begin{equation}
    {\rm mag}_{b, \rm deredden} = {\rm mag}_{b} -  a_{b,V} E(B-V)(\alpha, \delta),
\end{equation}
where ${\rm mag}$(${\rm mag}_{\rm deredden}$) are the observed and dereddened magnitudes in the corresponding band $b$.   $\alpha, \delta$ represent the right ascension and declination. 

\begin{table}
\small
\centering
\caption{
Component surveys (names, types, confidence, incremental number of objects (total cross-match without duplicates and confidence cuts), and reference) of the redshift reference sample.
Redshift type: s = spec-$z$, g = grism-$z$, p = multiband photo-$z$.\\
 {* Note: multiband photo-$z$ redshifts and grism and spec-$z$s with confidence
< 0.90 were not used in any training sets employed in this note.}\\
{{**Original datasets in this catalog are }\citet{2016ApJS..225...27M,2012MNRAS.425.2116C,2018A&A...619A.147P,2004A&A...421..913W,2006A&A...449..951G,2009AJ....138.1022S,2008AJ....135.1624S,2002ApJ...571..218N,2008ApJ...682..985W,2001MNRAS.322L..29C,2004ApJ...601L...5V,2004ApJ...600L.127D,2007ApJ...669..776K,2006ApJ...636..115P,2005ApJ...626..666M,2008ApJ...673..686H,2009ApJ...697..942R,2005A&A...437..883M,2005MNRAS.361..525D,2004A&A...428.1043L,2015ApJS..218...15K,2015AJ....149..178M,2009A&A...494..443P,2007A&A...465.1099R,2004ApJS..155..271S,2013ApJ...763....6T,2015ApJ...811...26T,2018A&A...616A.174P,2018MNRAS.479...25M,2010A&A...512A..12B,2009ApJ...695.1163V,2015A&A...576A..79L,2013A&A...559A..14L,2008A&A...478...83V,2009ApJ...706..885W,2010ApJS..191..124S}}
}
\begin{tabular}{p{1.5cm} p{0.5cm} p{0.5cm} p{1.5cm} p{2.8cm}}
\hline
Survey & Type & Conf & \#  & Ref. \\
\hline
\hline
2dFGRS & s & 1.00 & 3 (8) & \citet{colless2001} \\
       &   & 0.99 & 4 & \\
       &   & 0.90 & 1 & \\
2dflens & s & 1.00 & 1  & \citet{blake2016} \\
2MRS & s & 0.95 & 1 (6) & \citet{huchra2012} \\
6dFGRS & s & 0.98 & 2 & \citet{jones2009} \\
3D-HST & g & 0.99 & 5 (3803) & \citet{momcheva2016} \\
       &   & 0.95 & 277 & \\
ASTRODEEP & s & 1.00 & 4165 (13861) & \citet{merlin2021} ** \\
          & p$^{*}$ & 0.97 & 8212 & \\
ASTRODEEP-JWST & s & 1.00 & 594 (6303) & \citet{merlin2024} \\
               & p$^{*}$ & 0.92 & 628 & \\
               &   & 0.90 & 455 & \\
CANDELS & s & 1.00 & 53 (13447) & \citet{kodra2023} \\
        & p$^{*}$ & 0.93 & 6 & \\
JADES & s & 0.99 & 11 (318) & \citet{deugenio2025} \\
      &   & 0.95 & 34 & \\
      &   & 0.90 & 24 & \\
MOSDEF & s & 0.99 & 9 (45) & \citet{kriek2015} \\
NED & s & 0.95 & 847 (2956) & \citet{helou1991} \\
OzDES & s & 0.99 & 897 (910) & \citet{lidman2020} \\
PRIMUS & g$^{*}$ & 0.92 & 3653 (6263) & \citet{cool2013} \\
       &   & 0.85 & 1687 & \\
VANDELS & s & 1.00 & 196 (414) & \citet{garilli2021} \\
VIMOS & s & 1.00 & 499 (1343) & \citet{balestra2010} \\
      &   & 0.95 & 43 & \\
VUDS & s & 1.00 & 9 (150) & \citet{tasca2017} \\
     &   & 0.95 & 9 & \\
     &   & 0.80 & 3 & \\
VVDS & s & 1.00 & 101 (656) & \citet{lefevre2013} \\
     &   & 0.95 & 193 & \\
\hline
Totals & s & & 7699 & \\
      & g & & 5622 & \\
      & p & & 9301 & \\
      & all & & 22622 & \\
\hline
\end{tabular}
\label{tab:reference_catalogs}
\end{table}

\subsection{Reference Sample}
\label{sec:data:reference}

To construct the photometric redshift training and testing sets, we assembled a reference sample in the Extended Chandra Deep Field South (ECDFS) by collecting galaxies with spec-$z$s, grism-$z$s, and high-quality multiband photo-$z$ from multiple surveys listed in Table~\ref{tab:reference_catalogs}.

% The redshift reference catalog was then cross-matched to the LSSTComCam DP1 catalog using a radius of $0.75''$.

Confidence, which takes values between 0.0 and 1.0, is loosely defined as the probability that an individual redshift estimate is correct.
Most of the spectroscopic sets provide these estimates for their redshifts.
For the few that do not, we assigned the confidence as 0.95.
For the grism and multiband photo-$z$ surveys,  { we set the confidence equal to $1 - f_\text{out}$, where $f_\text{out}$ is the reported outlier rate of these catalogs.}
To facilitate custom quality cuts, the catalog contains flags indicating whether each redshift originates from spectroscopy, grism, or multiband photo-$z$, as well as confidence values.

The component redshift catalogs were combined into a single reference catalog.
When combining the component redshift catalogs, sources within $0.75''$ were identified as duplicates.
For these sources, only the highest quality redshift is kept, i.e., spec-$z$s are preferred over grism-$z$s, which are preferred over photo-$z$, and higher confidence values are preferred for redshifts of the same type.

Note that redshifts from grism and photo-$z$ surveys have larger scatter and bias than spectroscopic surveys; however, these characteristics are not captured by the confidence parameter.  { We also note the redshift quality in some grism surveys, e.g., PRIMUS, is similar to multi-band photo-$z$ due to limited wavelength resolution  \citep{cool2013}.}
We encourage users of our catalog to investigate the details of each component survey that comprises our reference catalog and apply their own quality cuts as suit their needs.

\begin{figure*}
    \centering
    \includegraphics[width=1.0\linewidth]{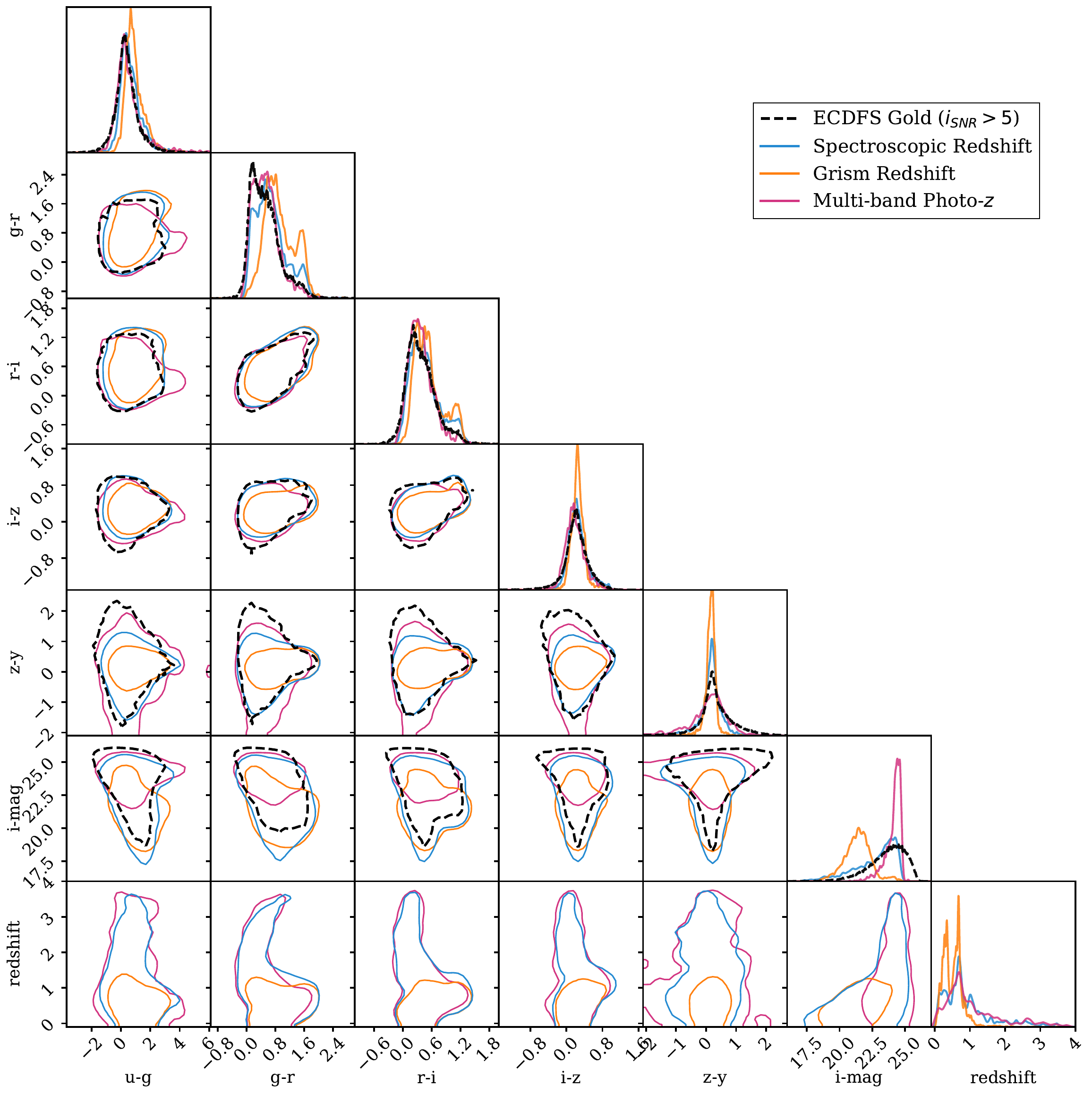}
    \caption{Corner plot of the training set color-magnitude-redshift distributions in the \texttt{ECDFS} field. Each panel shows \cwr{$95\%$} confidence contours among redshift, adjacent-band colors ($u-g$, $g-r$, $r-i$, $i-z$, $z-y$), and the $i$-band magnitude. The blue contours represent galaxies with spec-$z$s, the green contours represent grism-$z$s, the red contours represent the multiband photo-$z$, { while the black contours show all galaxies in the \texttt{ECDFS} field with i-band SNR greater than 5.} The spec-$z$, grism-$z$ and photo-$z$ samples compose $53\%$, $34\%$, and $13\%$ of the entire training/testing galaxies, respectively.  {Our reference sample covers a reasonable color space of our \texttt{ECDFS} samples; however, in some color spaces, we still lack reference galaxies.}  }
    \label{fig:color_space}
\end{figure*}

\subsubsection{Training and Testing Sets}
\label{sec:data:train}

We cross-matched the aforementioned reference galaxies to the \texttt{ECDFS} DP1 object catalog using a 0.75 \arcsec\ radius.  {To ensure sufficiently high quality photometry, we define the training set as all cross-matched objects with detections in all observing bands and with an $i$-band CModel flux SNR ratio exceeding 20. We note that galaxy model fluxes have underestimated errors, so the true SNR of the cut is likely lower than 20. We also note that requiring detections in all 6 bands will likely eliminate all ``dropout'' galaxies in the catalog, e.g., Lyman-break galaxies \citep{Giavalisco2002}. }

We also require the ``confidence'' to be greater than $0.9$.  We then split the selected sample, with 70\% used for training and the remaining 30\% reserved for a test set.  These selections result in sets of 6778 training galaxies and 2905 test galaxies.  {The average confidence of the selected sample is $0.97$, indicating approximately $\sim 3\%$ redshift outliers in the training set.} Note that we tolerate this relatively high level of redshift outliers due to the limited sample size for the training galaxies. 

 {In Fig.~\ref{fig:color_space}, we show the color-magnitude-redshift space of the training galaxies, and compare that to the color-magnitude space of all \texttt{ECDFS} galaxies with i-band SNR$>5$. With the combination of spec-$z$, grism-$z$ and multiband photo-$z$, the training set covers most color space of the \texttt{ECDFS} sample. Noticeably, the magnitude of the training galaxies only goes down to $i_{\rm mag} < 24.6$, compared to $i_{\rm mag} < 26.7$ for the object catalog. }

We note that we are training our photo-$z$ algorithms on training galaxies with higher significance than the object catalog. Only high-significance photometry is taken to reduce noise in the training process.  { Future study is required to determine the optimal selection criteria for maximizing performance on a faint galaxy catalog. }
Fig.~\ref{fig:train_info} provides an overview of the \texttt{ECDFS} training and testing sets.

 {We note that the PRIMUS survey is dominating the low-$z$ range, and the astrodeep sample is dominating the high-$z$ range. Potentially, the selection criteria of these specific surveys might be imprinted onto our photo-$z$. }

\subsubsection{DESI DR1}
\label{sec:data:desi}

\begin{figure}
    \centering
    \includegraphics[width=1\linewidth]{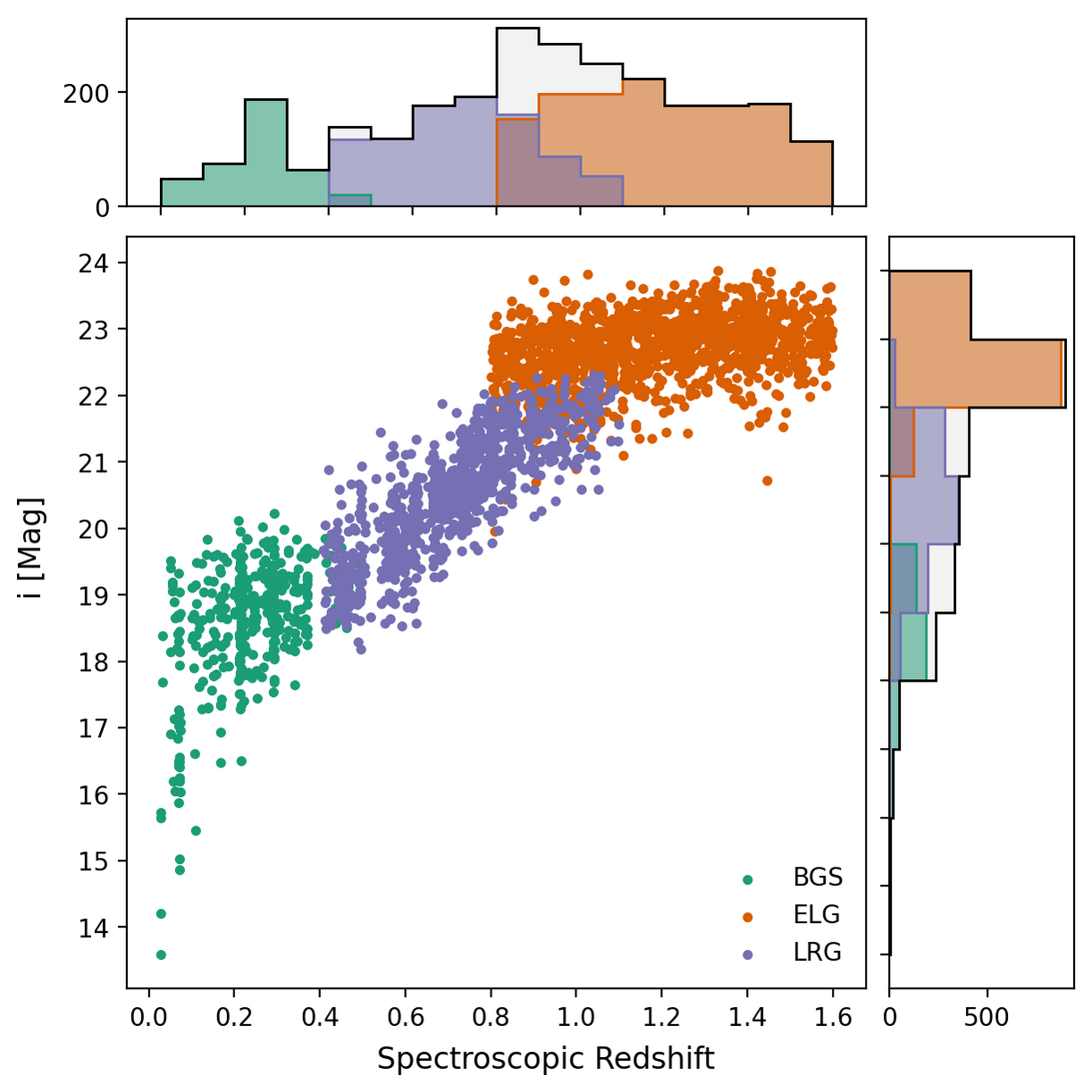}
    \caption{Redshift and $i$-magnitude distribution for matched DESI objects. For the scatter plot and histograms, the BGS sample is shown in green, ELG in orange, and LRG in purple. The distribution of the total sample is shown in black in the outer histograms.  { The $i$-mag distribution peaks at $i=23$ and the redshift distribution at $z = 0.8$.}}
    \label{fig:desi_redshift}
\end{figure}

For an independent validation of the four-band photometric redshift estimates, we cross-matched the \texttt{SV\_38\_7} DP1 object catalog with galaxies from the DESI Bright Galaxy Sample ( {BGS}), Luminous Red Galaxy (LRG), and Emission Line Galaxy (ELG) samples from the DESI DR1 spectroscopic catalog \citep{DESI_DR1}. This cross-match produced 2,728 matched objects across $z=0$ to $1.6$. 
We used these galaxies as a validation set to assess the performance of the four-band photo-$z$ estimates in a field with shallower depth and fewer filters. 
The $i$-band magnitudes and matched spec-$z$s of the galaxies are shown in Fig.~\ref{fig:desi_redshift}.

Matching was performed purely using spatial coordinates. If a match was found in multiple subcatalogs, the one with the highest weight was selected.  
The BGS sample, which provides redshifts for $z \leq 0.5$, contains no sources with declination $\delta \geq 7^\circ$, and covers only half of the \texttt{SV\_38\_7} field, limiting our ability to assess the low-redshift photo-$z$.  
Furthermore, in the matched catalog, the maximum $5\sigma \,i$-band magnitude depth is $23.8$, with a magnitude distribution peak at $i = 22.9$—both approximately 1.5 magnitudes shallower than the photometry for the full field.  
Despite these limitations, this sample provides an independent spectroscopic validation set and a test to apply the model trained on deep-field training catalogs to a wide field photometry.

\subsubsection{Euclid Crossmatch}
\label{sec:data:euclid}

To obtain joint optical and NIR photometry for galaxies in the \texttt{ECDFS} field, we positionally cross-matched the DP1 catalog with the Euclid Quick Data Release (Q1) \citep{Euclid2025} catalog. 
The Euclid catalog was filtered to retain only sources with a SNR ratio (SNR) greater than 5 in the VIS band and with no quality flags raised. For each matched source, we retained Euclid coordinates ($\alpha$, $\delta$), the point-spread function (PSF) magnitudes in the VIS band as well as uniform aperture magnitudes in the Y, J, and H NIR bands. All Euclid fluxes were converted to magnitudes using a zero point magnitude of  {$23.9$}. Sources with flagged measurements in any of the Y, J, or H bands (flag\_{band} = 1) were excluded from those bands by setting their magnitudes and uncertainties to NaN. 

 {The $5\sigma$ limit magnitudes of the Y, J, H bands from Euclid Q1 are around 24.0, and the VIS band magnitude limit is 26. The Euclid infrared magnitude limit is about 2 mag shallower than the $ugriz$ band of ECDFS. However, since we are applying an SNR cut above 20 to the i-band for the training and testing galaxies, most training and testing galaxies have a detection in the Euclid YJH bands. }

To study the benefit of including NIR photometry from Euclid, we created a training and testing set with the Euclid Y, J, and H photometry in addition to the DP1 $u, g, r, i, z, y$ photometry. We dropped all non-detections in Y, J, or H bands in addition to the selection made in Section~\ref{sec:data:train}. This results in $5010$ training set galaxies and $2158$ testing set galaxies, a $\sim 26\%$ decrease compared to the count of previously selected DP1 training and testing galaxy counts.

\section{Methodology}
\label{sec:method:0}

In this section, we briefly describe the eight photo-$z$ algorithms used in this work. We refer readers to \citet{RAIL} and \url{https://rail-hub.readthedocs.io/} for details about RAIL and these algorithms.

\begin{figure*}
    \centering
    \includegraphics[width=1.0\linewidth]{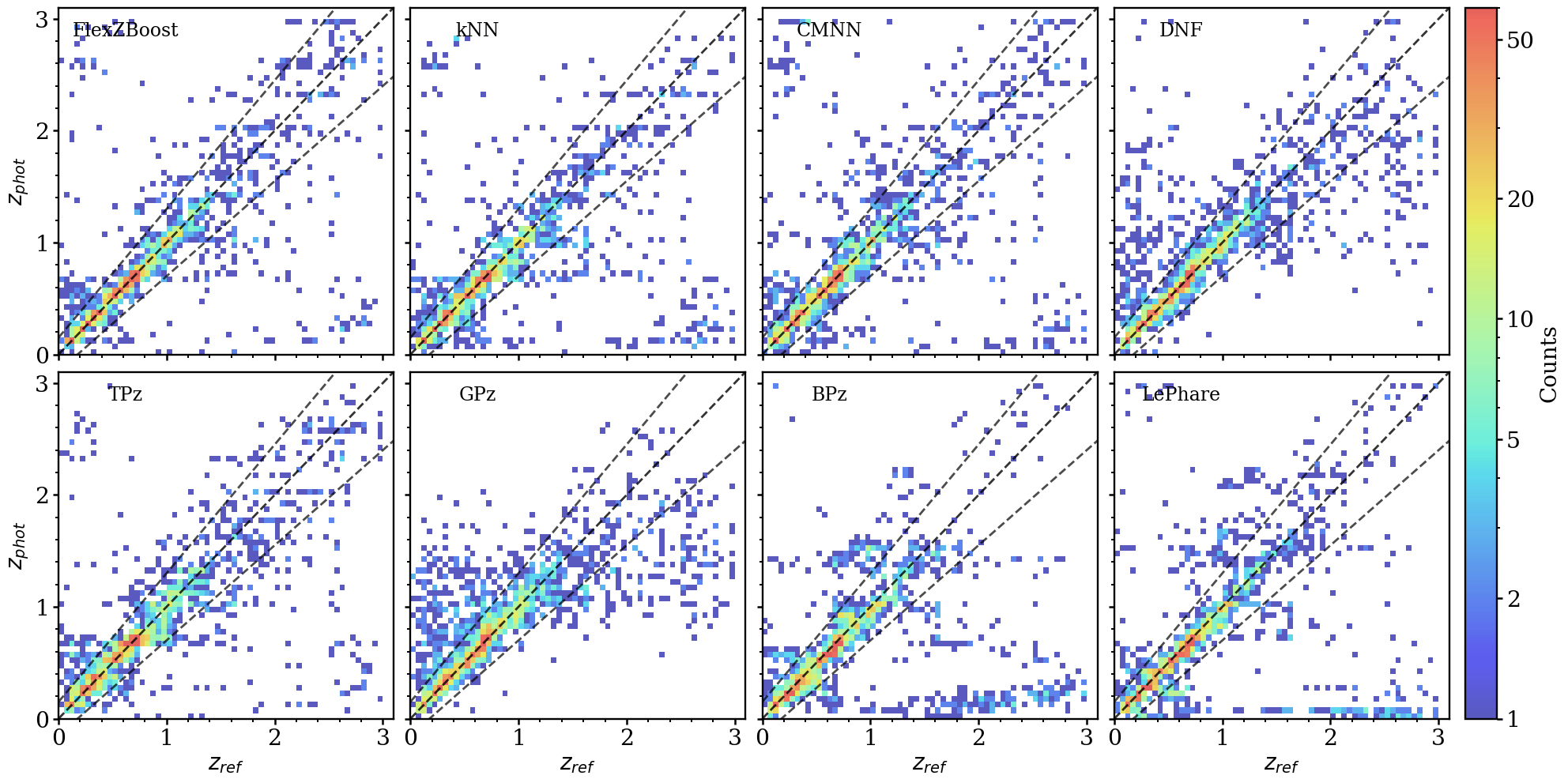}
    \caption{Two-dimensional histograms comparing the mode of the photometric redshifts ($z_{\mathrm{phot}}$) to reference redshifts ($z_{\mathrm{ref}}$) for eight photo-$z$ algorithms applied to the DP1 test sample. Each panel shows one algorithm: \texttt{FlexZBoost}, \texttt{kNN}, \texttt{CMNN}, \texttt{DNF}, \texttt{TPZ},  \texttt{GPz}, \texttt{BPZ}, and \texttt{LePhare}. The grey dashed lines indicate the identity line ($z_{\mathrm{phot}}=z_{\mathrm{ref}}$) and the $|\Delta z| = 0.15$ lines.   { The color scale represents the number of objects in each bin on a logarithmic scale. These plots illustrate the overall agreement and outlier behavior of each algorithm across the redshift range $0 < z < 3$.} }
    \label{fig:scatter}
\end{figure*}

\subsection{Template-Fitting Photo-$z$ Algorithms}
\label{sec:method:template}

Template-fitting photo-$z$ algorithms estimate redshifts by comparing observed galaxy photometry to a library of spectral energy distribution (SED) templates spanning a range of galaxy types and redshifts.  In this study, we deploy two widely used template-fitting codes: \texttt{BPZ} \citep{Benitez:2000,Coe:06} and \texttt{LePhare} \citep{Arnouts:1999,Ilbert:09,Ilbert2024}. Both codes compute synthetic fluxes by redshifting the templates and convolving them with the Rubin filter responses to produce model fluxes, then calculating $\chi^2$ values by comparing these model fluxes to the observed photometry and uncertainties for each template at each position on a sample redshift grid. These $\chi^2$ values are converted to likelihoods, and Bayesian priors are applied to incorporate expected redshift distributions  {as a function of magnitude and/or galaxy type.}   {A one-dimensional redshift posterior probability is computed by marginalizing (summing) over the template SEDs to produce the final PDF.} The best-fit redshift is identified as the mode of the posterior probability, and the best-fit SED is defined as the SED with the maximum contribution to the posterior at that single fixed redshift, and is not valid at other redshifts, and thus, not a precise representation of the galaxy-type vs. redshift degeneracy.

Both codes use a base SED set described in \citet[]{Ilbert:09} (also included in the base \texttt{LePhare} distribution), consisting of Elliptical and Spiral templates from \citet[]{Polletta:07} as well as bluer starburst SEDs generated from \citet[]{BruzualCharlot:03} models.  These base templates contain no internal dust extinction.  The \texttt{LePhare} code is designed to add a grid of E(B-V) values to each template with values \texttt{0.05,0.1,0.15,0.2,0.25,0.3,0.35,0.4,0.5}.  \texttt{BPZ} lacks this capability, and so new templates are created explicitly by adding extinction from \citet[]{Noll:09} (Calzetti-like dust with a UV bump) to each SED bluer than the \texttt{Sb} type (i.~e.~no dust is added to the Elliptical, S0, and Sa models, but dust is added to all other SEDs). E(B-V) extinctions with values \texttt{0.1,0.2,0.3,0.4,0.5} are used to generate the final set of 136 SEDs for BPZ.

As template-fitting methods can compute model fluxes at any redshift (as long as the SED model spans the wavelengths of the filter curves at that redshift), they can extrapolate beyond the limitations of sparse or missing redshifts in referece training sets, and thus have the potential to be more robust at higher redshifts.  However, the redshift predictions are limited by the agreement between the assumed SED models and the real Universe: any mismatch in galaxy evolution as a function of lookback time, luminosity, and other variables will lead to biases in the redshift predictions.  So, careful construction of template sets and calibration using deep multiwavelength samples will be critical to the performance of template-based methods.

\subsection{Machine-Learning Photo-$z$ Algorithms}
\label{sec:method:machine_learning}

Machine-learning photo-$z$ algorithms predict photometric redshifts by learning empirical mappings from multiband photometry to the reference redshift, modeling the complex color–-redshift relationships. They can be less prone to systematic bias in the magnitude measurement because they are learning the magnitude-redshift mapping empirically. However, machine-learning methods tend to predict redshift based on training set information on the target set by acting as implicit priors on the color--redshift relation, and can generate biased results if the training galaxies are not representative of the full galaxy catalog. 

The methods employed in this work are:
\begin{itemize}
    \item \texttt{TPZ} (Trees for photo-$z$) uses random forests to perform regression of redshift as a function of multiband photometry \citep{Carrasco-Kind:2013}\footnote{\url{http://github.com/LSSTDESC/rail\_tpz}}. 
    \item \texttt{FlexZBoost} uses boosted decision trees to model the non-linear mapping from colors to redshift \citep{Izbicki:2017}\footnote{\url{http://github.com/LSSTDESC/rail\_flexzboost}}.
    \item \texttt{kNN} (K-Nearest Neighbors) predicts the redshift of a galaxy based on the average redshift of its closest neighbors in color-magnitude space.\footnote{\url{http://github.com/LSSTDESC/rail\_sklearn}}
    \item \texttt{CMNN} (Color-Matched Nearest Neighbors) improves upon basic \texttt{kNN} by weighting neighbors according to Mahalanobis distances in color space \citep{Graham:2018}\footnote{\url{http://github.com/LSSTDESC/rail\_cmnn}}.
    \item \texttt{GPz} (Gaussian Processes for photo-$z$) models the redshift-color relation with Gaussian processes \citep{Almosallam:2016}\footnote{\url{http://github.com/LSSTDESC/rail\_gpz\_v1}}. 
    \item \texttt{DNF} (Directional Neighborhood Fitting) fits local linear models around each galaxy using its nearest neighbors \citep{2016MNRAS.459.3078D}\footnote{\url{http://github.com/LSSTDESC/rail\_dnf}}.
\end{itemize}

\begin{figure*}
    \centering
    \includegraphics[width=1.0\linewidth]{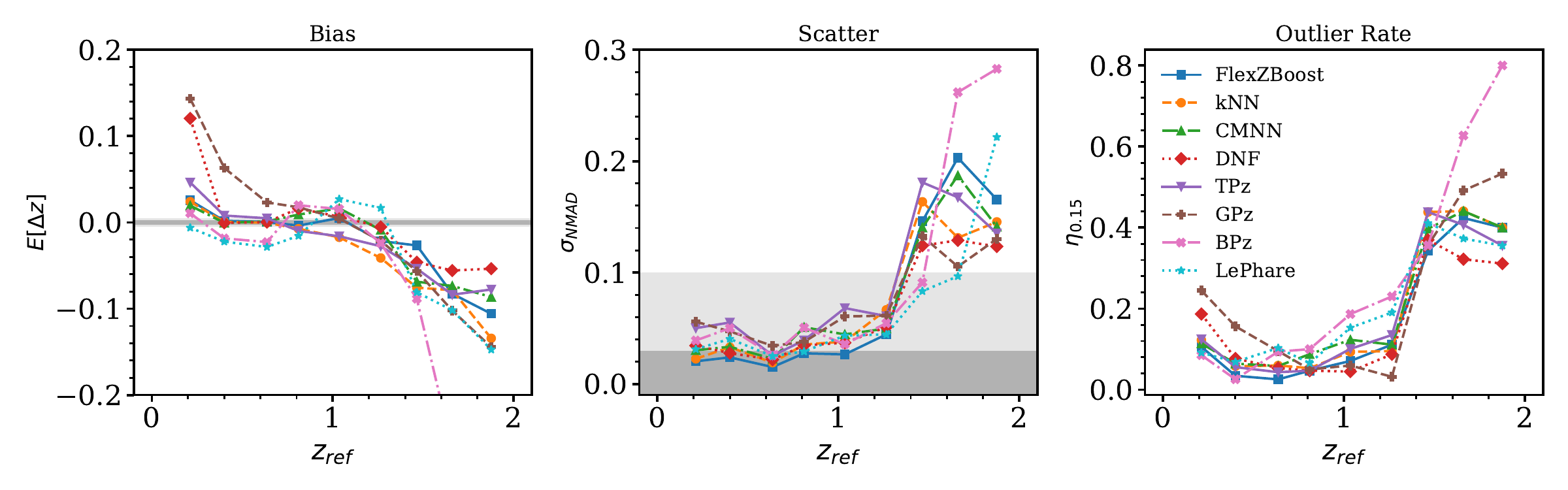}
    \includegraphics[width=1.0\linewidth]{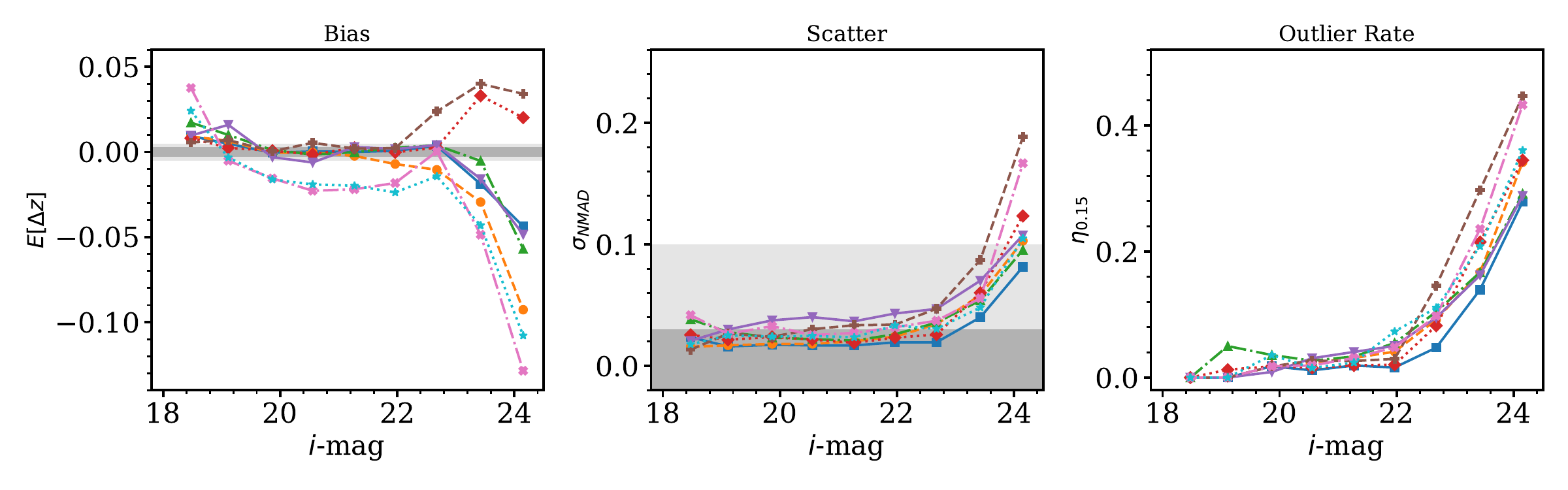}
    \caption{
    Photo-$z$ performance metrics as a function of redshift (top row) and $i$-band magnitude (bottom row) for eight algorithms.
    Each panel shows three metrics: the photo-$z$ bias $\mathbb{E} [ \Delta z ]$, scatter $\sigma_{\rm NMAD}$, and outlier rate $\eta_{0.15}$.
     { The grey shaded regions show the LSST Y1 and \cwr{Y10} requirements on the mean and scatter of photo-$z$ \citep{desc_srd}. The photo-$z$ of all algorithms starts to deteriorate when $z>1.2$ and $i$-mag$>23$. We notice that some algorithms display significant bias at low and high redshift, and template-fitting methods have more bias across the magnitude range than empirical methods. We do not show statistics beyond $z_{\rm ref} = 2$ because of the limited number of reference galaxies.  } }
    \label{fig:metrics}
\end{figure*}

\begin{figure}
    \centering
    \includegraphics[width=1\linewidth]{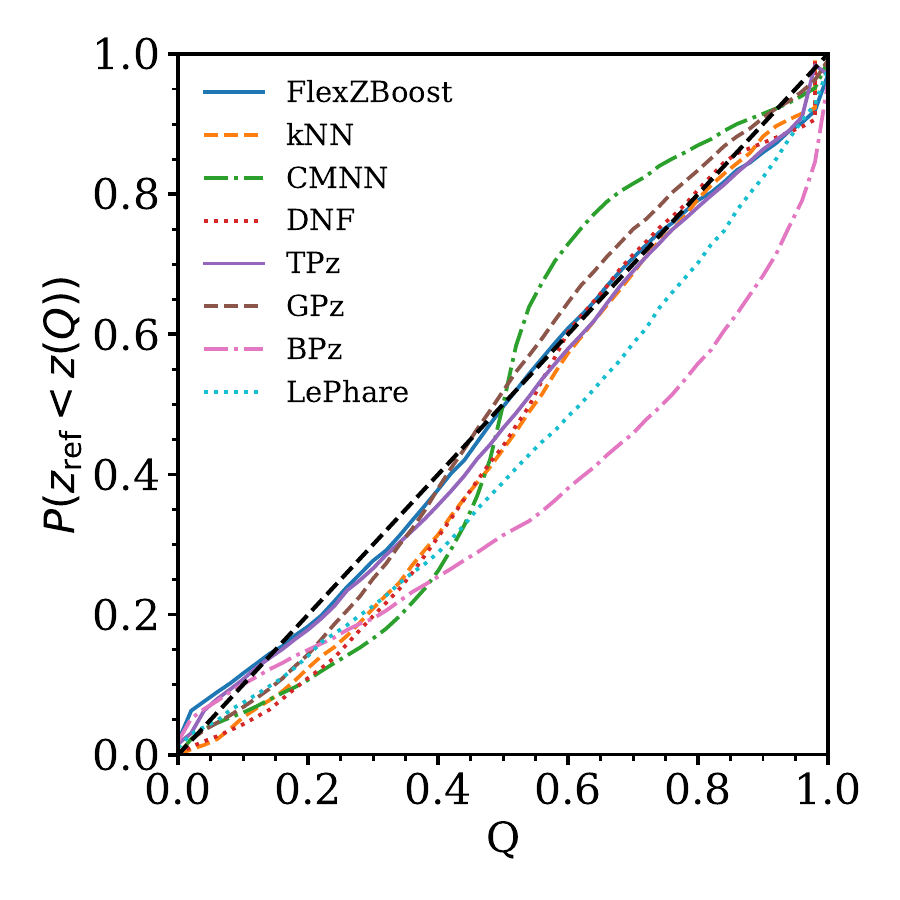}
    \caption{The PIT-QQ plot for eight photo-$z$ estimators. The curves show the empirical CDF of \(\{z_i\}\) as a function of \(Q\), i.e., \(P\!\big(z_{\mathrm{ref}}<z(Q)\big)\), where \(z(Q)\) is the \(Q\)-th posterior quantile. The black dashed line denotes perfect calibration (uniform PIT).}
    \label{fig:pit-qq}
\end{figure}

\subsection{Workflow Management and Bookkeeping}
\label{sec:method:rail_project}

We developed a workflow management and bookkeeping software system, \texttt{RAIL\_projects}\footnote{\url{https://github.com/LSSTDESC/rail\_projects}} for RAIL pipelines. The workflow manager produces configuration files for \texttt{ceci}\footnote{\url{https://github.com/LSSTDESC/ceci}}pipelines based on ``flavors'' that specify the input datasets, algorithms and corresponding parameters, and selection criteria. \texttt{RAIL\_projects} enables efficient comparisons between algorithms, parameters, and selection criteria, which can be highly useful in future LSST analyses and photo-$z$ analyses in other surveys. The configuration of the pipelines are stored in \texttt{RAIL\_project\_config}\footnote{\url{https://github.com/LSSTDESC/rail\_project\_config}}.

In this work, there are three flavors of photo-$z$ models being trained: 
\begin{enumerate}
    \item \textbf{dp1\_6band}: models trained on the $ugrizy$ Gaap1p0 magnitude of the training and testing galaxy sets in ECDFS. These models are then used to produce photo-$z$ of the objects in the ECDFS,  \texttt{EDFS} and \texttt{SV\_95\_-25} fields. This configuration can be found in \texttt{dp1/dp1\_v4.yaml}.
    \item \textbf{dp1\_4band}: models trained on the $griz$ Gaap1p0 magnitude of the training and testing galaxies in ECDFS. These models are used to produce photo-$z$ of the object in the  \texttt{SV\_38\_7} field.
    \item \textbf{dp1\_paper\_euclid\_nir}: only \texttt{FlexZBoost} and \texttt{BPz} are applied in this flavor, using the DP1 $ugrizy$ and Euclid $YJH$ magnitude. The model is only applied to the test galaxies in ECDFS. 
\end{enumerate}

\begin{figure*}
	\centering
	\includegraphics[width=1.0\linewidth]{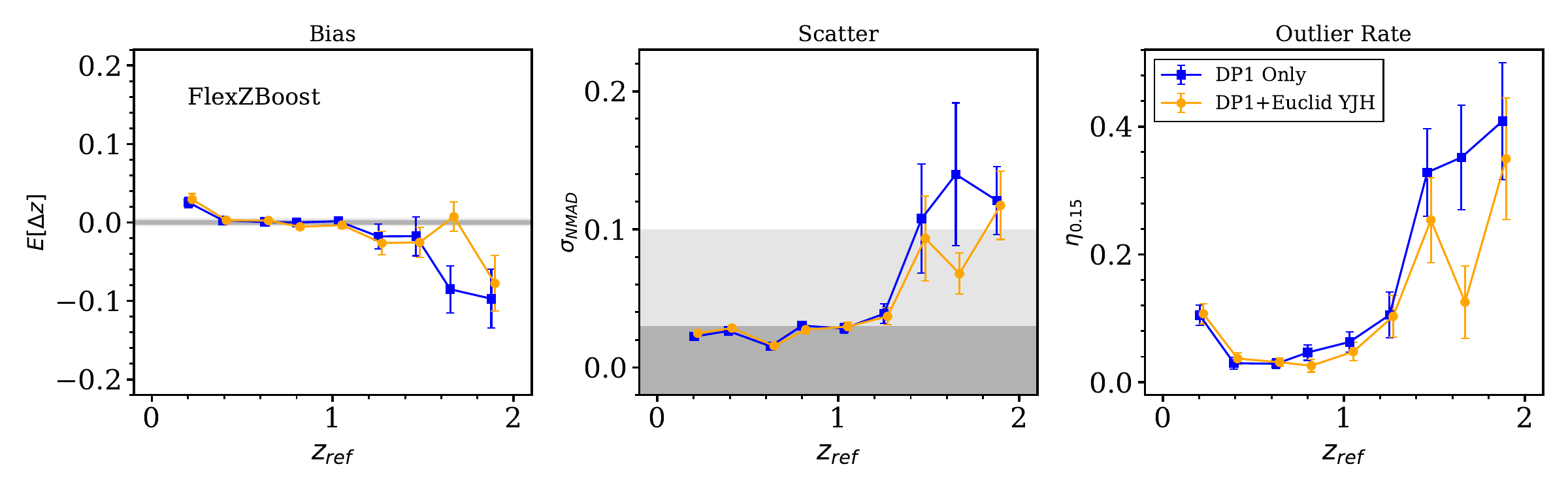}
    \caption{
    The same {photometric} redshift performance metrics as Fig.~\ref{fig:metrics} of the \texttt{FlexZBoost} algorithm as a function of redshift, comparing LSST ComCam-only photometry (blue) to the combination of LSST ComCam and Euclid NIR YJH bands (orange). 
     {The inclusion of Euclid NIR data notably reduces bias, scatter, and outlier rates at redshift $z>1.5$.} The error bars are computed by bootstrapping the testing set. \cwr{The grey shaded regions show the LSST Y1 and Y10 requirements on the mean and scatter of photo-$z$.}
    }
	\label{fig:dp1_euclid}
\end{figure*}

\subsection{Metrics}
\label{sec:method:metrics}

In this study, we use simple metrics based on the mode of the photo-$z$ and the reference redshift to evaluate the performance of the photo-$z$ algorithms. We compute the photo-$z$ deviation by
\begin{equation}
	\Delta z = \frac{z_{\rm mode} - z_{\rm ref}}{1+z_{\rm ref}}.
\end{equation}
Here $z_{\rm mode}$ is the redshift mode, i.e., the redshift that corresponds to the maximum of the PDF. $z_{\rm ref}$ is the reference redshift, sometimes referred to as the ``true redshift''. 

For a given set of galaxies, we compute the bias as the mean of $\Delta z$,  {the scatter as the bias-corrected normalized median absolute deviation (NMAD) $\sigma_{\rm NMAD}$ of $\Delta z$,}
\begin{equation}
\sigma_{\rm NMAD} = 1.48 \times {\rm median} \left( \left| \Delta z - {\rm median}(\Delta z)\right| \right),
\end{equation}
 {and the catastrophic outlier rate $\eta_{0.15}$ by the fraction of galaxies with $\Delta z > 0.15$. }

Photo-$z$ are often used to assign  {both lens and source galaxies to tomographic bins} in joint analyses of galaxy clustering and weak lensing \citep[e.g.,][]{des_y3_3x2pt}. Therefore, we compute a ``binning accuracy'' as a metric for binning performance, defined as the probability that the bin assigned based on the photo-$z$ mode is the same as that assigned using $z_{\rm ref}$. The binning strategy we use is to split all galaxies into five bins with equal number counts. We note that the binning accuracy should be taken as a qualitative metric to assess the accuracy of binning for the algorithm, rather than as a metric for assessing the impact on cosmology analysis. 

\cwr{We use the Probability Integral Transform quantile–quantile (PIT-QQ) plot as a diagnostic tool to evaluate the coverage of the photo-$z$ PDFs. By comparing the empirical cumulative density function (CDF) at quantile $Q$, i.e., $P(z_{\rm ref} < z(Q))$ against the diagonal line, the plot shows whether the predicted PDFs are well calibrated. Deviations from the diagonal line indicate systematic issues such as overconfident, underconfident error bars, biased, or skewed uncertainties.}

\section{Results}
\label{sec:res:0}

\begin{figure*}
	\centering
	\includegraphics[width=0.45\linewidth]{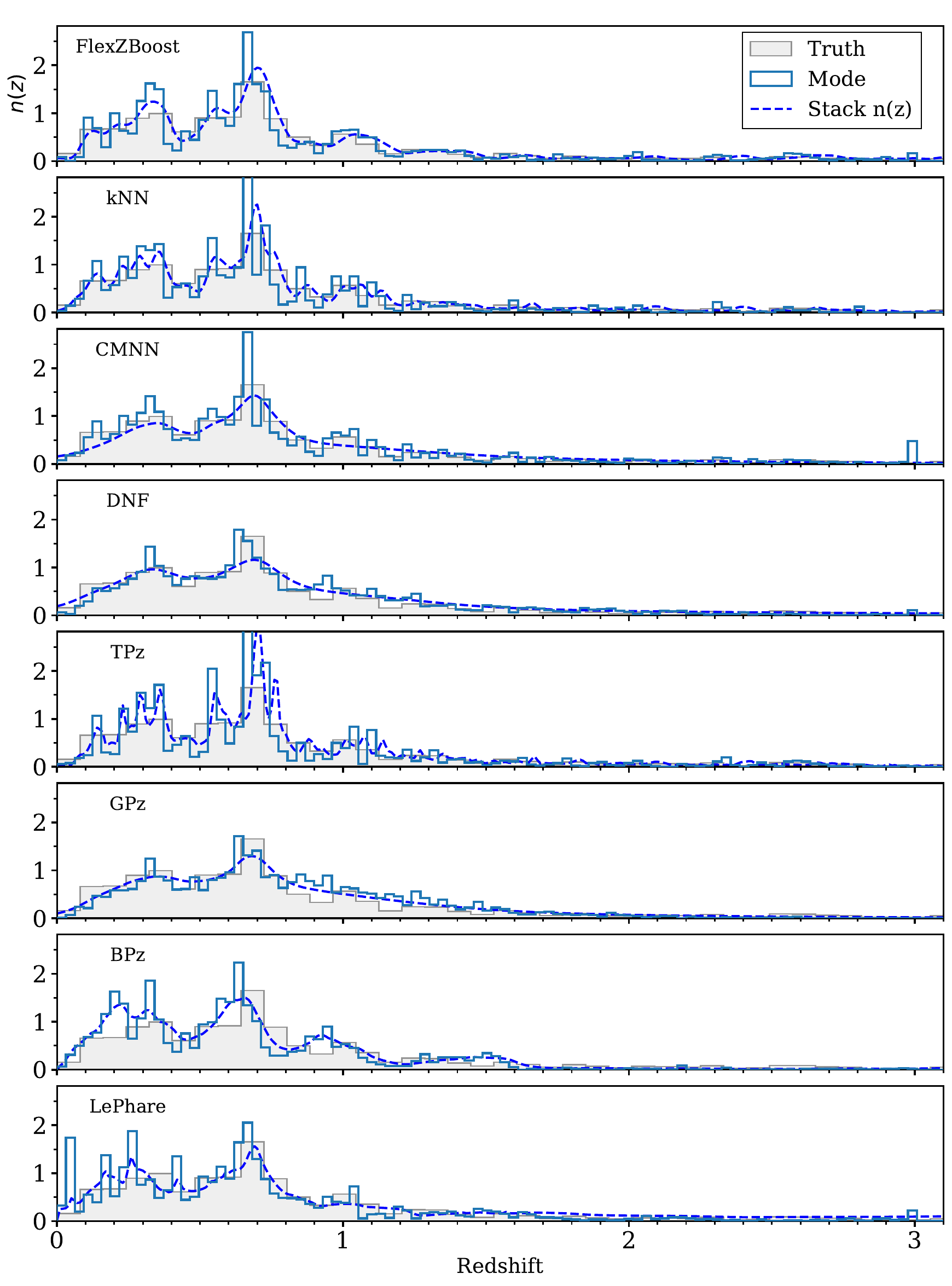}
	\includegraphics[width=0.45\linewidth]{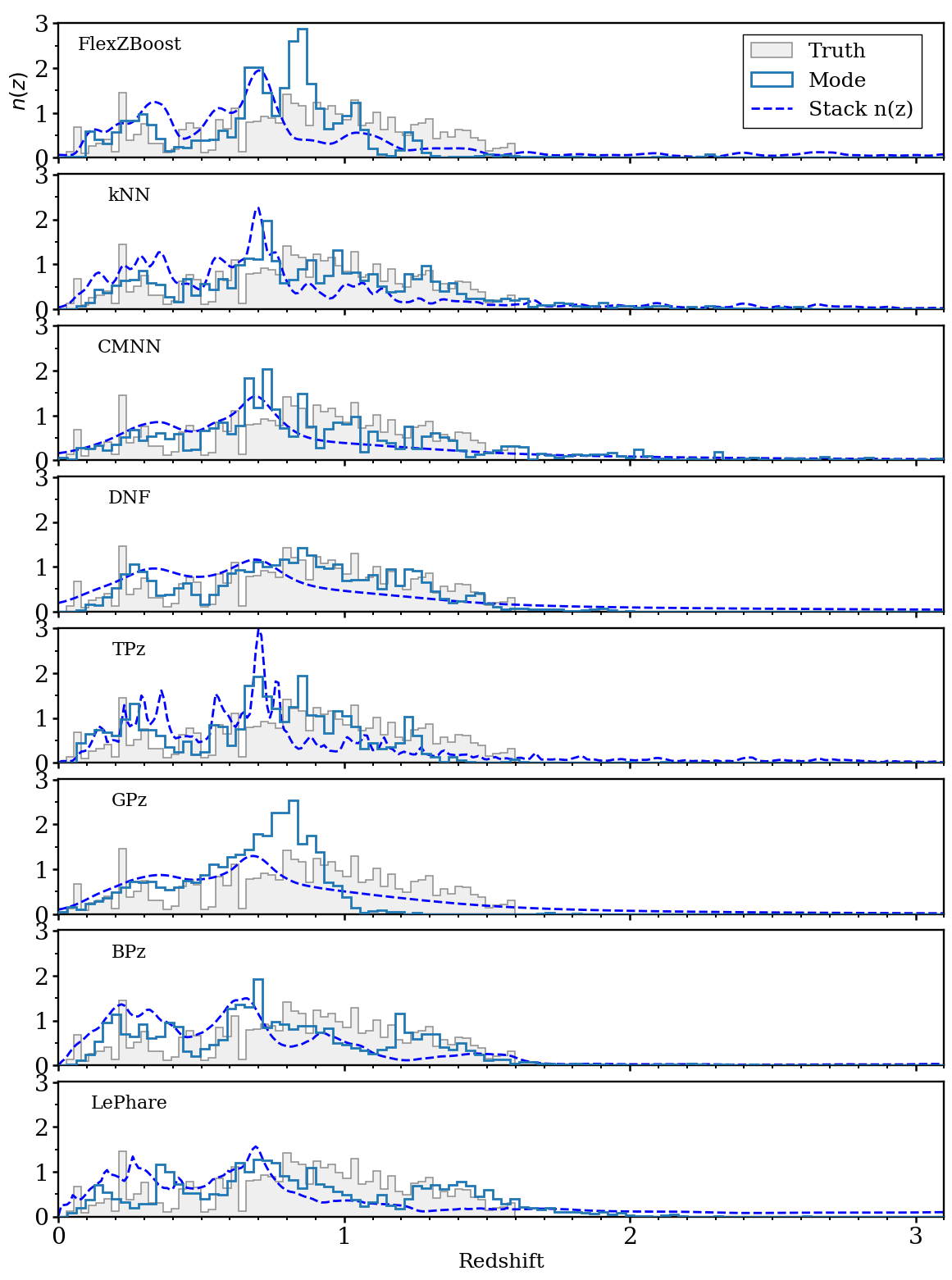}
	\caption{Comparisons of stacked redshift distributions estimated by each photo-$z$ algorithm with the true redshift distributions for the \texttt{ECDFS} test set (left panels) and the independent DESI cross-match in the \texttt{SV\_38\_7} field (right panels). Each panel shows, for one algorithm, the stacked redshift distribution $n(z)$ (blue dashed lines) estimated from the photometric redshift PDFs, the histogram of point-estimate photo-$z$ values (blue step lines), and the histogram of the true redshift distribution (grey distribution).  {These plots demonstrate the ability of each algorithm to recover the underlying redshift distribution. All profiles in this figure are normalized.} }
	\label{fig:nz}
\end{figure*}

\subsection{Photo-$z$ Optimization}

In this section, we briefly describe the effort towards achieving an optimized photo-$z$ result in this work. We note that we do not consider the results here to be the fully optimized photo-$z$ performance for future Rubin data. 

As an initial data exploration, we train and test all eight photo-$z$ algorithms with default settings using different types of photometry. We observe that the Gaap1p0, the CModel, and the Sersic flux are generally better performing than other types of flux measurements in terms of scatter and outlier rate metrics, and they are comparable to each other within minor fluctuations between algorithms. As a result, we choose Gaap1p0 as flux measure employed for the rest of the work.

We adjust the values of the hyperparameters of the machine-learning methods and adjust the template collections for the template-fitting algorithms to optimize the performance of the photo-z methods. Here are the notable hyperparameter changes compared to the default parameters in the \texttt{RAIL v1.2}\footnote{The full set of hyperparameter change can be found in the ``dp1\_paper'' flavor in \url{https://github.com/LSSTDESC/rail_project_config/blob/main/dp1/dp1.yaml}}:
\begin{itemize}
	\item Common settings: PDF and point estimates are calculated between $z=0$ and $3$.
	\item \texttt{TPz}: number of random bootstrap samples is set to 10  {(default is 8)}, number of trees in the random forest is set to 10  {(default is 5)}, the minimum number in terminal leaf is set to 2  {(default is 5)}.
	\item \texttt{kNN}: maximum number of neighbors is set to 10 (default is 7). 
	\item \texttt{GPz}: use the variable diagonal covariance mode. The number of training iterations is set to 1000  {(default is 200, and does not converge)}. 
    \item \texttt{FlexZBoost}: maximum number of basis functions is set to 50  {(default is 35)}.
	\item \texttt{BPz}:
    use the base templates from \texttt{LePhare} with a set of dust absorption models applied, resulting in a larger range of dusty model SEDs than standard \texttt{BPZ} runs (See Section~\ref{sec:method:template}). Use a Hubble Deep Field North (HDFN) like prior from \citet{Benitez:2000}, with no zero-point adjustments.
	\item \texttt{LePhare}: enable magnitude zero point adjustment (default is off). Remove a maximum of two bands when $\chi^2$ of the fit exceeds $300$  {(default is $500$)}.
\end{itemize}

\subsection{Per-galaxy Photo-$z$}
\label{sec:res:per_galaxy}

We evaluate the per-galaxy photo-$z$ performance of the eight algorithms on our test set, which has the same population as the training set. 
Fig.~\ref{fig:scatter} shows that most algorithms achieve good agreement with the spectroscopic redshifts for bright, low-to-moderate redshift galaxies, as seen by the concentration of points along with the identity line. 
However, systematic biases and increased scatter become apparent at higher redshifts ($z \gtrsim 1.2$), where training data are sparse and photometric uncertainties are high. 

The patterns of outlier distribution vary across algorithms.  {Particularly, for the template-fitting algorithms, we notice that there are galaxies with $1.5<z_{\rm ref}<3$ that are given $z_{\rm phot}<0.5$. We suspect this can be caused if the Lyman break of the galaxy are misidentified as a Balmer break by the algorithm. }

We summarize the overall performance metrics in Table~\ref{tab:performance}.  {Overall, the machine-learning based algorithms have bias less than $0.005$, which is LSST Y1 requirement \citep{desc_srd}. } Some algorithms, such as \texttt{FlexZBoost} and \texttt{DNF}, and \texttt{kNN} exhibit tighter scatter, while others, like GPz, show larger overall dispersion.   {All algorithms has scatter less than $0.1$, the LSST Y1 requirement. We also report the catastrophic outlier rates defined as the fraction of galaxies with  $|\Delta z| > 0.15$. }In terms of binning accuracy, \texttt{FlexZBoost} scores the highest accuracy of $78.1\%$, while other methods are in the range of $60\%$ to $70\%$.

\begin{table}
\centering
\caption{Performance metrics for photo-$z$ algorithms using 6-band DP1 data in the \texttt{ECDFS} field. Columns show the redshift bias $\mathbb{E}[\Delta z]$, the scatter measured by $\sigma_{\rm NMAD}$,  the outlier rate {$\eta_{\,0.15}$}, and the tomographic binning accuracy. }
\begin{tabular}{lcccc}
\hline
{Algorithm} & $\mathbb{E}[\Delta z]$ & \textbf{$\sigma_{\rm NMAD}$} & \textbf{$\eta_{\,0.15}$} & bin. acc.\\
\hline
 \texttt{FlexZBoost} & -0.0005 & 0.0280 &  0.117 & 0.781 \\
 \texttt{kNN}        & -0.0035 & 0.0379 &  0.149 & 0.689 \\
 \texttt{CMNN}       &  0.0001 & 0.0416 &  0.145 & 0.723 \\
 \texttt{DNF}        & -0.0001 & 0.0361 &  0.152 & 0.698 \\
 \texttt{TPZ}        &  0.0010 & 0.0547 &  0.138 & 0.689 \\
 \texttt{GPz}        &  0.0025 & 0.0551 &  0.204 & 0.634 \\
 \texttt{BPZ}        & -0.0179 & 0.0459 &  0.192 & 0.587 \\
 \texttt{LePhare}    & -0.0143 & 0.0400 &  0.171 & 0.628 \\
\hline
\end{tabular}
\label{tab:performance}
\end{table}

In Fig.~\ref{fig:metrics}, we show the bias, scatter, and outlier rate of the algorithms as functions of redshift and $i-$band magnitude. Generally, we observe that the photo-$z$ performance is better at intermediate redshift ($0.3<z<1.2$), and brighter magnitude ($i-$mag $<22.5$). We also observe a wide range of performance across different methods vs. redshift, e.g., the template-fitting methods tend to exhibit more significant bias.  {We also noticed that the bias for \texttt{BPz} and \texttt{DNF} are particularly significant at low redshift ($z_{\rm ref} < 0.5$), while \texttt{DNF} provides the lower bias, scatter, and outlier rate at the highest redshift. The bias of the photo-$z$ algorithms across the redshift exceeds the LSST Y1 requirement on the mean redshift \citep{desc_srd}, highlighting the importance of redshift calibration via techniques like clustering redshift \citep{Newman08}. The scatter of the photo-$z$ are within the LSST Y1 requirement for $z_{\rm ref} < 1.2$. } 

The variety of performance at different redshift and magnitude ranges means the optimal photo-$z$ algorithm depends on the science cases one wants to explore. This is a particular advantage of RAIL, as it is relatively easy to compute photo-$z$ by multiple methods. 

\cwr{In Fig.~\ref{fig:pit-qq}, we show the PIT-QQ curves of all methods we are evaluating. The general trend of the empirical CDF follows the identity lines. However, we note that some methods are giving overconfident error bars, e.g., CMNN. We also notice the template fitting methods \texttt{Bpz} and \texttt{LePhare} have PDF skewed toward high redshift, thus a PIT-QQ curve lower than the diagonal line. }

In addition, we also train models using four-band photometry ($griz$), and estimate the photo-$z$ on the \texttt{SV\_38\_7} field. We also evaluate performance for galaxy ensembles binned by redshift and magnitude. We find that each of our photo-$z$ algorithms performs well in the range of $z=0.2$ to $1.2$. We refer readers to \citet{SITCOMTN-154} for discussion of the performance with magnitude and redshift, and details about the four-band photo-$z$.

\subsection{photo-$z$ Including Euclid NIR Photometry}
\label{sec:res:euclid}

We study the impact on the photo-$z$ performance of using galaxies with DP1 + Euclid NIR photometry, described in Section~\ref{sec:data:euclid}.  {For simplicity}, we only used the \texttt{FlexZBoost} algorithm, which was trained and tested on this dataset both with and without the Euclid NIR photometry. 

In Fig.~\ref{fig:dp1_euclid}, we show the comparisons of the bias, scatter, and outlier rate as a function of redshift for \texttt{FlexZBoost} with/without Euclid NIR photometry. Since our train/test galaxies concentrate at redshift $z<1.2$, the infrared information does not significantly improve the photo-$z$ bias and scatter overall.  {We found considerable improvement in the bias, scatter and outlier rate for test galaxies at $z_{\rm ref}>1.2$.}
To study the full improvement of including near-infrared photometry in LSST photometry, future studies need to obtain more high-redshift reference samples to reduce the uncertainties of the metrics.

\subsection{Redshift Distribution Estimation}
\label{sec:res:per_galaxy}

In this section, we compare the redshift distributions predicted by each photo-$z$ algorithm against the true redshift distributions from the reference sample. We evaluate both stacked $n(z)$ estimates, obtained by summing per-galaxy photometric redshift PDFs, and the distributions of point estimates. These comparisons enable us to evaluate the ability of each algorithm to recover the true redshift distribution, which is essential for robust weak lensing and clustering measurements for DESC.

Fig.~\ref{fig:nz} shows the stacked $n(z)$ and the mode histogram compared to the true $n(z)$ distribution for each algorithm. For the \texttt{ECDFS} test set with six-band photometry, most photo-$z$ algorithms' stacked $n(z)$ successfully recover the true redshift distribution. This suggests that, given deep multiband data and a representative training set, both machine-learning and template-fitting methods can provide reliable ensemble redshift estimates.  {Again, we notice that the template-fitting algorithms have a peak at redshift $z_{\rm ref}\sim0.2$, which may be caused by misidentifying Lyman break galaxies as low redshift Balmer break galaxies. }

In the four-band \texttt{SV\_38\_7} validation field, some machine-learning methods exhibit imprints of the training set redshift distribution, producing tails at high redshift that the true redshift distribution does not have {, as well as a spike at redshift $z\sim0.7$.} Fundamentally, this is caused because the \texttt{ECDFS} training galaxies and \texttt{SV\_38\_7} galaxies have significantly different color-magnitude-redshift distribution.  Of all the methods, we find that the mode histogram of \texttt{DNF} best matches the true distribution. Several of the stacked $n(z)$ distribution exhibit significant systematics that warrant further investigation, e.g., the stacked $n(z)$ of \texttt{CMNN}, \texttt{GPz}, and \texttt{LePhare}. \cwr{Similar spikes can also be seen in the stacked PDF of these methods, shown in Appendix~\ref{ap:stacked_pz}.}

 {We note that all training and testing galaxies are drawn from a relatively small field. Therefore, our photo-$z$ are subject to training set sample variance \citep{Cunha2012} when applying the photo-$z$ algorithm to a different field, e.g., \texttt{SV\_38\_7}. }

\section{Conclusion}
\label{sec:conclu:0}

In this work, we use the Rubin Observatory's DP1 dataset to demonstrate the capability of RAIL to produce a photometric redshift for LSST and DESC. These productss include per-galaxy PDF, their point estimates, and ensemble $n(z)$, using various machine-learning and template-fitting algorithms. 

We construct a reference set in the \texttt{ECDFS} and cross-match to DP1 galaxies, and cross-match the DP1 galaxies to Euclid and DESI to obtain infrared photometry and external validation redshifts. The cross-matched galaxies are used as training and testing galaxies for the machine-learning models, and calibration galaxies for the template-fitting methods. 

We have demonstrated that the algorithms in RAIL can produce photo-$z$ results with promising accuracy and precision for real LSST data. With a representative training set, the biases of the machine-learning algorithms are less than $0.005$, which is the LSST Y1 requirement \citep{desc_srd} for overall redshift bias. The photo-$z$ scatter can reach $\sim 0.03$ level,  {below the LSST Y1 requirement of $0.1$}, and the outlier rate can reach $\sim 10\%$ for a high SNR sample.  We study the improvement of photo-$z$ by including the Euclid NIR photometry, and find minor improvement for photo-$z$ bias and outlier rate at high redshift. We demonstrate that the point estimate of the photo-$z$ can be used for tomographic binning, with \texttt{FlexZBoost} reaching $0.78$ binning accuracy. 
The stacked $n(z)$ shows good agreement with the true distribution in the \texttt{ECDFS} test set. For the \texttt{SV\_38\_7} field, we find that the \texttt{DNF} point estimates result in the least biased $n(z)$, while some methods display a significant level of systematic error. 

The results we present have several notable caveats that are worth noting: 
The size and depth of the training set limit the robustness of the photo-$z$ past a redshift of $z\approx1.2$. 
Some photo-$z$ algorithms perform poorly with non-detections in some bands. 
We undertook limited effort in optimizing the hyperparameters of the algorithms, so additional performance gains likely could be found in future work. 
The red leak, i.e., non-zero infrared transmission, of the LSSTComCam g-band might impact the g-band photometry and the photo-$z$. We do not expect this issue to arise in LSSTCam.

% \tianqing{add another caveat about g-band red leakage, and that we do not expect it to be an issue for LSSTCam}

We identify several potential future improvements. To name a few: (a) non-detection/negative fluxes in a subset of bands need to be better handled by a few algorithms; (b) algorithm-specific quality flags should be calculated and recorded; (c) combining different flux measurements may provide additional flux information; this will be explored in the future; (d) training set choices can be further examined to maximize performance on the galaxies in Rubin's object catalog. 

In summary, this work lays a solid foundation for applying RAIL to real LSST datasets. The DP1 photo-$z$ results show promising results for future analysis. The work demonstrates the readiness of the pipeline, but it also highlights some areas that require future progress. 

% % Example table
% \begin{table}
% 	\centering
% 	\caption{We can show some metrics here}
% 	\label{tab:example_table}
% 	\begin{tabular}{lccr} % four columns, alignment for each
% 		\hline
% 		Methods/Metrics & TPZ & C & D\\
% 		\hline
% 		1 & 2 & 3 & 4\\
% 		2 & 4 & 6 & 8\\
% 		3 & 5 & 7 & 9\\
% 		\hline
% 	\end{tabular}
% \end{table}

\section*{Contribution Statements}

T. Zhang: analysis and coordination, reference catalog construction, algorithm optimization, developer of RAIL, writing the manuscript.  \\
E. Charles: core developer of RAIL and \texttt{rail\_projects}, photo-$z$ data production, project oversight. \\
J.F. Crenshaw: construction of the reference sample and Euclid cross-match. \\
S. Schmidt: core developer of RAIL, algorithm optimization, and paper editing, and project coordination. \\
P. Adari: construction of the DESI cross-match and analysis. \\
J. Gschwend: developer of the PZ server.\\
S. Mau: developer of \texttt{rail\_projects}. \\

\section*{Acknowledgements}

TZ, AM, DO, OL are supported by Schmidt Sciences. TZ thanks SLAC National Accelerator Laboratory for providing hospitality and an excellent research environment during the course of this study.
JFC acknowledges support from the U.S. Department of Energy, Office of Science, Office of High Energy Physics Cosmic Frontier Research program under Award Number DE-SC0011665. 
This research is supported by the U.S. Department of Energy under awards DE-SC0025309 and DE-SC0023387 received by PA. 
This research received support from the National Science Foundation (NSF) under grant No. NSF DGE-1656518 through the NSF Graduate Research Fellowship received by SM. 
MJJ acknowledges support for the current research from the National Research Foundation (NRF) of Korea under the programs 2022R1A2C1003130 and RS-2023-00219959. 
SL is supported by the U.S. Department of Energy under grant number DE-1161130-116-SDDTA and under Contract No. DE-AC02-76SF00515 with the SLAC National Accelerator Laboratory. 
The work of AAPM was supported by the U.S. Department of Energy under contract number DEAC02-76SF00515.  AAPM thanks the Department of Physics and the Laboratory of Particle Astrophysics and Cosmology at Harvard University, the Cosmology Group at Boston University, and the Department of Physics at Washington University in St. Louis. 
EU is supported by the U.S. Department of Energy and the Cosmic Frontier grant DE-SC0007881. 
AvdL is supported by the US Department of Energy under award under award DE-SC0025309. 
CWW was supported by Department of Energy, grant DE-SC0010007.

The photo-$z$ Server uses computational resources of IDAC-Brazil at the Laboratório Interinstitucional de e-Astronomia (LIneA) with financial support from INCT do e-Universo (Process no. 465376/2014-2).

ISN, LTSC and JdV are partially supported by Spanish Ministerio de Ciencia, Innovacion y Universidades under grant PID2021-123012. ISN, LTSC and JdV gratefully acknowledges funding from the MAD4SPACE-CM TEC-2024/TEC-182 project funded by Comunidad de Madrid.

This material is based upon work supported in part by the National Science Foundation through Cooperative Agreements AST-1258333 and AST-2241526 and Cooperative Support Agreements AST-1202910 and AST-2211468 managed by the Association of Universities for Research in Astronomy (AURA), and the Department of Energy under Contract No.\ DE-AC02-76SF00515 with the SLAC National Accelerator Laboratory managed by Stanford University.
Additional Rubin Observatory funding comes from private donations, grants to universities, and in-kind support from LSST-DA Institutional Members.

 {
This paper has undergone internal review in the LSST Dark Energy Science Collaboration. 
The internal reviewers were Qianjun Hang and Brett Andrews. We thank them for their valuable contribution. We thank Yihao Zhou and Lei Hu for their helpful input on the paper. }

The DESC acknowledges ongoing support from the Institut National de 
Physique Nucl\'eaire et de Physique des Particules in France; the 
Science \& Technology Facilities Council in the United Kingdom; and the
Department of Energy and the LSST Discovery Alliance
in the United States.  DESC uses resources of the IN2P3 
Computing Center (CC-IN2P3--Lyon/Villeurbanne - France) funded by the 
Centre National de la Recherche Scientifique; the National Energy 
Research Scientific Computing Center, a DOE Office of Science User 
Facility supported by the Office of Science of the U.S.\ Department of
Energy under Contract No.\ DE-AC02-05CH11231; STFC DiRAC HPC Facilities, 
funded by UK BEIS National E-infrastructure capital grants; and the UK 
particle physics grid, supported by the GridPP Collaboration.  This 
work was performed in part under DOE Contract DE-AC02-76SF00515.

%%%%%%%%%%%%%%%%%%%%%%%%%%%%%%%%%%%%%%%%%%%%%%%%%%
\section*{Data Availability}
\label{sec:data_avail}

The DP1 photo-z data products can be accessed in several ways.

\subsection*{Rubin Science Platform}

DP1 photo-$z$ data can be accessed at the Rubin Science Platform\footnote{\url{data.lsst.cloud}}. 
During the DP1 commissioning, we have demonstrated functionality to a) ingest model files produced by training the various algorithms in the Rubin Data Butler (Butler), b) use the Rubin Data Management software stack along with photo-$z$ specific plugins \footnote{\url{https://github.com/lsst-dm/meas_pz}} to produce pre-object PDF estimates in the data management framework, with the estimates automatically being written to the Butler,  c) retrieve those $p(z)$ estimates as we would retrieve any other Rubin data product.  Work is ongoing to fully support this functionality in Rubin Data Preview 2 (DP2), with the estimates for the \texttt{kNN} and \texttt{BPZ} algorithms being supported and distributed by the Rubin Observatory for DP2, and other algorithms being supported and distributed by the Rubin science community.

% add a link to dp1 tutorial 

\subsection*{Photo-$z$ Server}

The Photo-z Server \footnote{\url{https://pzserver.linea.org.br}} is a web-based service available for the LSST community to create and host lightweight PZ-related data products. All data products described in this document will be hosted on the Photo-z Server, along with their respective metadata and documentation. A list with access instructions and links to product pages will be available on the data product documentation page \footnote{\url{https://docs.linea.org.br/en/data/pz\_server\_data.html\#data-preview-1}}.

\subsection*{LSDB}

The Large Scale DataBase (LSDB) \citep{LSDB} hosts DP1 data as well as the photo-$z$ point estimates from this work at \url{https://data.lsdb.io/#Rubin_DP1/object_photoz}.

% \brycek{Maybe DESC DP1 Website will have value-added catalogs publicly available?}

%%%%%%%%%%%%%%%%%%%% REFERENCES %%%%%%%%%%%%%%%%%%

% The best way to enter references is to use BibTeX:

\bibliographystyle{mnras}
\bibliography{main} % if your bibtex file is called example.bib

\bigskip
\hrule
\bigskip
\bigskip

\noindent $^{1}$Department of Physics and Astronomy and PITT PACC, University of Pittsburgh, Pittsburgh, PA 15260, USA\\
$^{2}$SLAC National Accelerator Laboratory, 2575 Sand Hill Road, Menlo Park, CA 94025, USA\\
$^{3}$Kavli Institute for Particle Astrophysics and Cosmology (KIPAC), Stanford University, Stanford, CA 94305, USA\\
$^{4}$Department of Physics, University of Washington, Seattle, WA 98195, USA\\
$^{5}$DIRAC Institute, University of Washington, Seattle, WA 98195, USA\\
$^{6}$Department of Physics and Astronomy, University of California, One Shields Avenue, Davis, CA 95616, USA\\
$^{7}$Department of Physics and Astronomy, Stony Brook University, Stony Brook, NY 11794-3800, USA\\
$^{8}$Laboratório Interinstitucional de e-Astronomia - LIneA, Rua Gal. José Cristino 77, Rio de Janeiro, RJ - 20921-400, Brazil\\
$^{9}$Université Paris Cité, CNRS, Astroparticule et Cosmologie, F-75013 Paris, France\\
$^{10}$Department of Physics and Astronomy, San José State University, San José, CA 95192-0106, USA\\
$^{11}$Department of Physics, University of Wisconsin-Madison, Madison, WI 53706, USA\\
$^{12}$Laboratoire d'Annecy de Physique des Particules (LAPP), 9 Chem. de Bellevue, B.P. 110, 74941 Annecy, France\\
$^{13}$Laboratoire de Physique des 2 Infinis Irène Joliot-Curie (IJClab), Bât. 100, 15 rue Georges Clémenceau, 91405 Orsay, France\\
$^{14}$Ohio University, Athens, OH 45701, USA\\
$^{15}$Université Clermont-Auvergne, CNRS, LPCA, 63000 Clermont-Ferrand, France\\
$^{16}$LPSC-IN2P3, Laboratoire de Physique Subatomique et Cosmologie, Universit'e Grenoble-Alpes, CNRS/IN2P3, 53, rue des Martyrs,38026 Grenoble Cedex, France\\
$^{17}$Steward Observatory, The University of Arizona, 933 N. Cherry Ave., Tucson, AZ 85721, USA\\
$^{18}$NSF–DOE Vera C. Rubin Observatory, 950 N. Cherry Ave., Tucson, AZ 85719, USA\\
$^{19}$Laboratoire de Physique Nucléaire et des Hautes Énergies (LPNHE), Sorbonne Université, Université Paris Cité, CNRS/IN2P3, F-75005 Paris, France\\
$^{20}$Centro de Investigaciones Energéticas, Medioambientales y Tecnológicas (CIEMAT), Madrid, Spain\\
$^{21}$Department of Physics and Astronomy, Rutgers University, Piscataway, NJ 08854, USA\\
$^{22}$Vera C. Rubin Observatory, Avenida Juan Cisternas \#1500, La Serena, Chile\\
$^{23}$Department of Physics \& Astronomy, University College London, Gower Street, London WC1E 6BT, UK\\
$^{24}$Laboratoire d'Astrophysique, Marseille Aix-Marseille University, 38 Rue Frédéric Joliot Curie, 13013 Marseille, France\\
$^{25}$University of Pennsylvania, Philadelphia, PA 19104, USA\\
$^{26}$Department of Physics, Yonsei University, 50 Yonsei-ro, Seodaemun-gu, Seoul 03722, South Korea\\
$^{27}$Departments of Physics and Astronomy, University of California, Berkeley, CA, USA\\
$^{28}$Department of Physics, Duke University, Durham, NC 27708, USA\\
$^{29}$Department of Astrophysical Sciences, Princeton University, Princeton, NJ 08544, USA\\
$^{30}$McWilliams Center for Cosmology and Astrophysics, Department of Physics, Carnegie Mellon University, Pittsburgh, PA, USA\\
$^{31}$Sidrat Research, 124 Merton Street, Suite 507, Toronto, ON M4S 2Z2, Canada\\
$^{32}$Infrared Processing and Analysis Center (IPAC), California Institute of Technology, 1200 E California Blvd, Pasadena, CA 91125, USA\\
$^{33}$Max-Planck-Institut für extraterrestrische Physik, Giessenbachstr. 1, 85748 Garching, Germany\\
$^{34}$Centre de Physique des Particules de Marseille, 163, avenue de Luminy Case 902 13288 Marseille cedex 09, France\\
$^{35}$Department of Physics, Harvard University, Cambridge, MA, USA\\
$^{36}$Institute of Astronomy, University of Edinburgh, Edinburgh EH9 3HJ, United Kingdom\\

\appendix

\section{Stacked PDF of all galaxies}
\label{ap:stacked_pz}

\begin{figure*}
	\centering
	\includegraphics[width=1.0\linewidth]{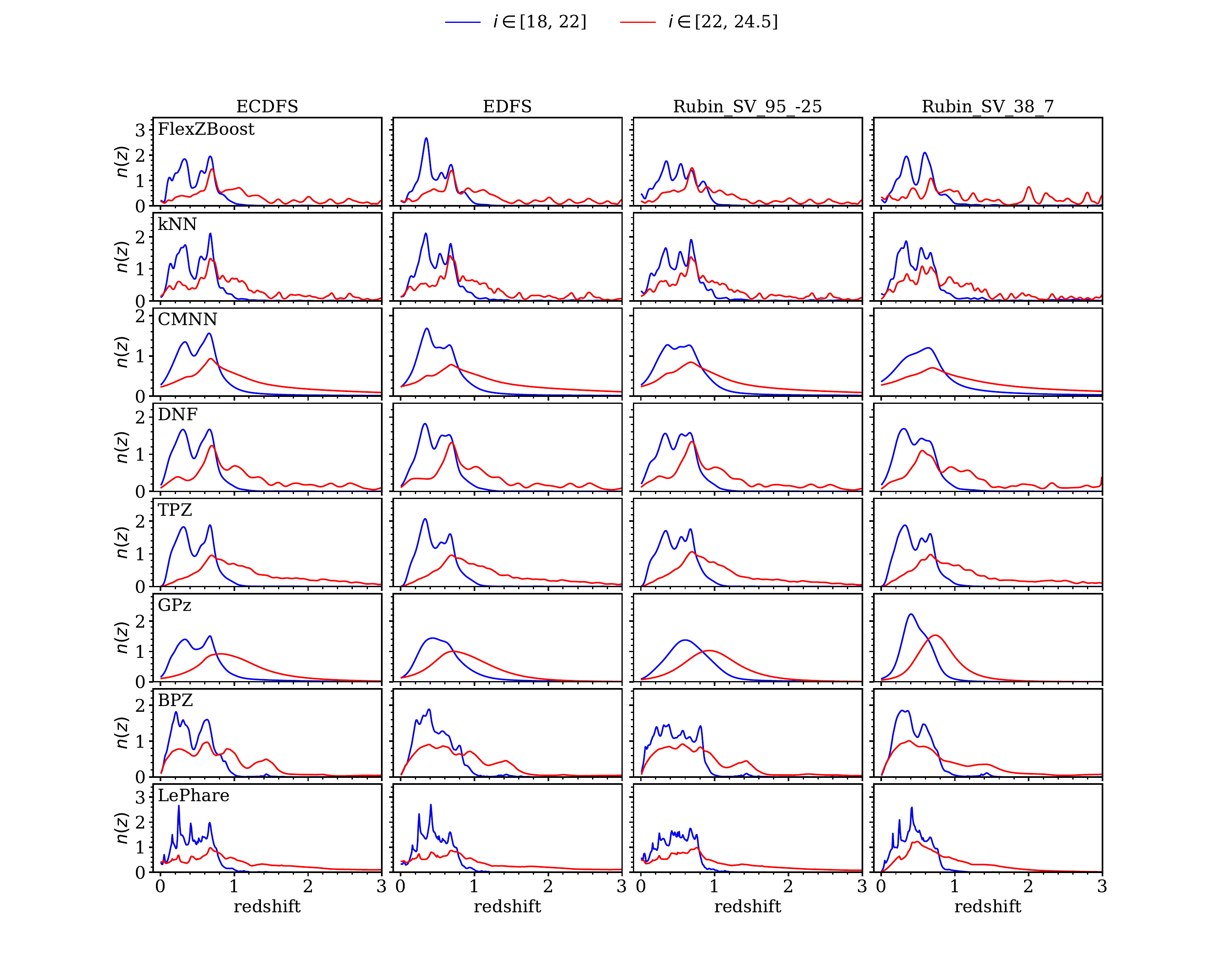}
    \caption{Stacked PDF of all DP1 gold samples in the four fields with multi-band observation, from all eight of our algorithms. The galaxies are binned into a bright bin ($i \in [18,22]$, blue lines) and a faint ($i \in [22, 24.5]$, red lines). The stacked PDFs of machine learning methods show the imprint of the training sample. 
    }
	\label{fig:stacked_pz}
\end{figure*}

\cwr{In Figure~\ref{fig:stacked_pz}, we show the stacked PDF of all galaxies in all four DP1 fields, binned by their $i$-band magnitude. The faint galaxies have overall higher redshifts, which follows the empirical expectation of a magnitude-limited sample.  We can see signs of overfitting for \texttt{FlexZBoost} and \texttt{kNN} at high redshift, as those peaks match across the fields and are likely not real. The double peak feature that appeared in our training set is imprinted onto most of the stacked PDFs. This highlights the importance of smoothing out the training set by reweighting and resampling in the future. }

% Don't change these lines
\bsp	% typesetting comment
\label{lastpage}
\end{document}